\documentclass[12pt]{article}

\usepackage{newtxtext,newtxmath}

\usepackage{graphicx}

\usepackage[letterpaper,margin=1in]{geometry}

\renewenvironment{abstract}
	{\quotation}
	{\endquotation}

\date{}

\makeatletter
\renewcommand{\fnum@figure}{\textbf{Figure \thefigure}}
\renewcommand{\fnum@table}{\textbf{Table \thetable}}
\makeatother

\usepackage{scicite}

\usepackage{url}

\newcommand{\dHs}{\Delta H_s}

\def\scititle{
	Poroelastic aquifer response drives seasonal vertical land motion in southern Louisiana
}
\title{\bfseries \boldmath \scititle}

\author{
	Pritom~Sarma$^{1,2\ast}$,
	Eduardo~Arzabala$^{2}$,
	Carolina~Hurtado-Pulido$^{6}$,
	Einat~Aharonov$^{1,3}$\and
	Renaud~Toussaint$^{4,3}$,
	Stanislav~Parez$^{5}$,
	Cynthia~Ebinger$^{2}$\and
	\small$^{1}$Institute of Earth Sciences, Hebrew University of Jerusalem, Jerusalem, Israel.\and
	\small$^{2}$Department of Earth and Environmental Sciences, Tulane University, New Orleans, LA, USA.\and
	\small$^{3}$PoreLab, The Njord Centre, Departments of Physics and Geosciences, University of Oslo, Oslo, Norway.\and
	\small$^{4}$Universit\'{e} de Strasbourg, CNRS, ENGEES, Institut Terre et Environnement de Strasbourg, UMR 7063, \and 
    \small Strasbourg, France.\and
	\small$^{5}$Faculty of Science, Jan Evangelista Purkyn\v{e} University in \'{U}st\'{i} nad Labem, \'{U}st\'{i} nad Labem, Czech Republic.\and
	\small$^{6}$Department of Earth, Atmospheric, and Planetary Sciences, Purdue University, West Lafayette, IN, USA.\and
	\small$^\ast$Corresponding author. Email: pritom.sarma@mail.huji.ac.il
}

\begin{document}

% Insert the title and author list
\maketitle

% Abstract -- 123 words (Science limit: 125). In bold.
% NB: Science forbids citations AND abbreviations in the abstract, so "GNSS" is
% not used here; it is defined at first use in the main text instead.
\begin{abstract} \bfseries \boldmath
Coastal Louisiana is sinking, amplifying flooding and land loss, yet the seasonal component of this motion resists attribution.
Satellite geodetic records at Baton Rouge from 2004 to 2024 reveal long-term subsidence of $-2.69\pm0.69$~mm\,yr$^{-1}$ carrying a superposed 10 to 15~mm annual oscillation, in phase with river stage and confined-aquifer head.
Such positive correlation is diagnostic of poroelastic rather than surface loading.
A poroelastic model of a semi-confined aquifer driven by head variations reproduces both signals, and the seasonal amplitude decays logarithmically with distance from the Baton Rouge Fault--Mississippi River intersection, as expected for radial diffusion from a flux source.
Fault--river intersections thus act as seasonal conduits into deep aquifers, an underappreciated control on coastal land motion that will grow with hydrological extremes.
\end{abstract}

% The first paragraph of any Science paper does NOT have a heading, nor is it indented
\noindent
Southern Louisiana sits on the Gulf of Mexico passive margin, where east--west-striking Mesozoic--Recent normal faults cut a thick wedge of Cenozoic sediments and host the metropolitan area of Baton Rouge $\sim$100~km inland from the coast~\cite{mcculloh2013surface,Hopkins2021}.
Vertical land motion here is the net result of several processes acting over different spatiotemporal scales: Holocene sediment compaction~\cite{Tornqvist2008,Jankowski2017,Song2025}, fault creep~\cite{Dokka2006,Shen2017,hurtado2024variations}, isostatic adjustment to past and present loads, and groundwater dynamics~\cite{Yuill2009,Karegar2015,hurtado2026quantifying}.
Because relative sea-level rise rates along this coast (1.4--13.2~mm\,yr$^{-1}$;~\cite{Kolker2011,Dangendorf2024}) far exceed the global mean~\cite{Nerem2018}, disentangling these processes is essential for hazard assessment.

Hydrological processes deform the solid Earth through two distinct mechanisms~\cite{vanDam2001,Argus2014,Larochelle2022}.
Surface (elastic) loading: surface and shallow water mass depresses the crust, producing an \textit{anti-correlation} between variations in the load and vertical ground motion~\cite{Farrell1972}.
Subsurface (poroelastic) response: pore-pressure changes within an aquifer drive volumetric strain that, for recharge, lifts the surface: a \textit{positive correlation} between hydraulic head and uplift, opposite in sign to elastic loading~\cite{Puskas2017,Carlson2024}.
Poroelastic displacements can exceed elastic loading by one to two orders of magnitude at short wavelengths and often dominate near large aquifer systems~\cite{hurtado2026quantifying}.
Separating the two is therefore a prerequisite for inverting geodetic data for water storage and for removing hydrological contamination from tectonic signals.

The Baton Rouge area provides an exceptional natural laboratory: the Mississippi River crosses two sub-parallel growth faults (the Baton Rouge and Denham Springs faults) where deep Pleistocene aquifers~\cite{tomaszewski2002water,Sargent2011} are juxtaposed against a major surface-water hydraulic source. Here we combine two decades of continuous GNSS, groundwater, and river-stage data with a simple analytical poroelastic model to show that (i) seasonal and decadal vertical land motion in this region are poroelastic in origin, (ii) the spatial pattern of seasonal motion uniquely identifies the fault--river intersection as a recharge source obeying a flux boundary condition, and (iii) climate-driven amplification of hydrological extremes will magnify this signal.
This question carries added urgency because the geologically young, diagenetically immature sedimentary formations of coastal Texas and Louisiana have been identified as among the most favorable settings in North America for large-scale geologic storage of CO\textsubscript{2}, both in saline aquifers and in depleted hydrocarbon reservoirs~\cite{Meckel2021,Bump2021,DeAngelo2019,Zulqarnain2023,ZobackHennings2025}.
The same fault--aquifer hydraulic connectivity we document below is therefore directly relevant to the long-term integrity of any such storage operation in the region.

\subsection*{Seasonal and decadal geodetic signals}

\begin{figure}[htbp]
\centering
\includegraphics[width=0.6\textwidth]{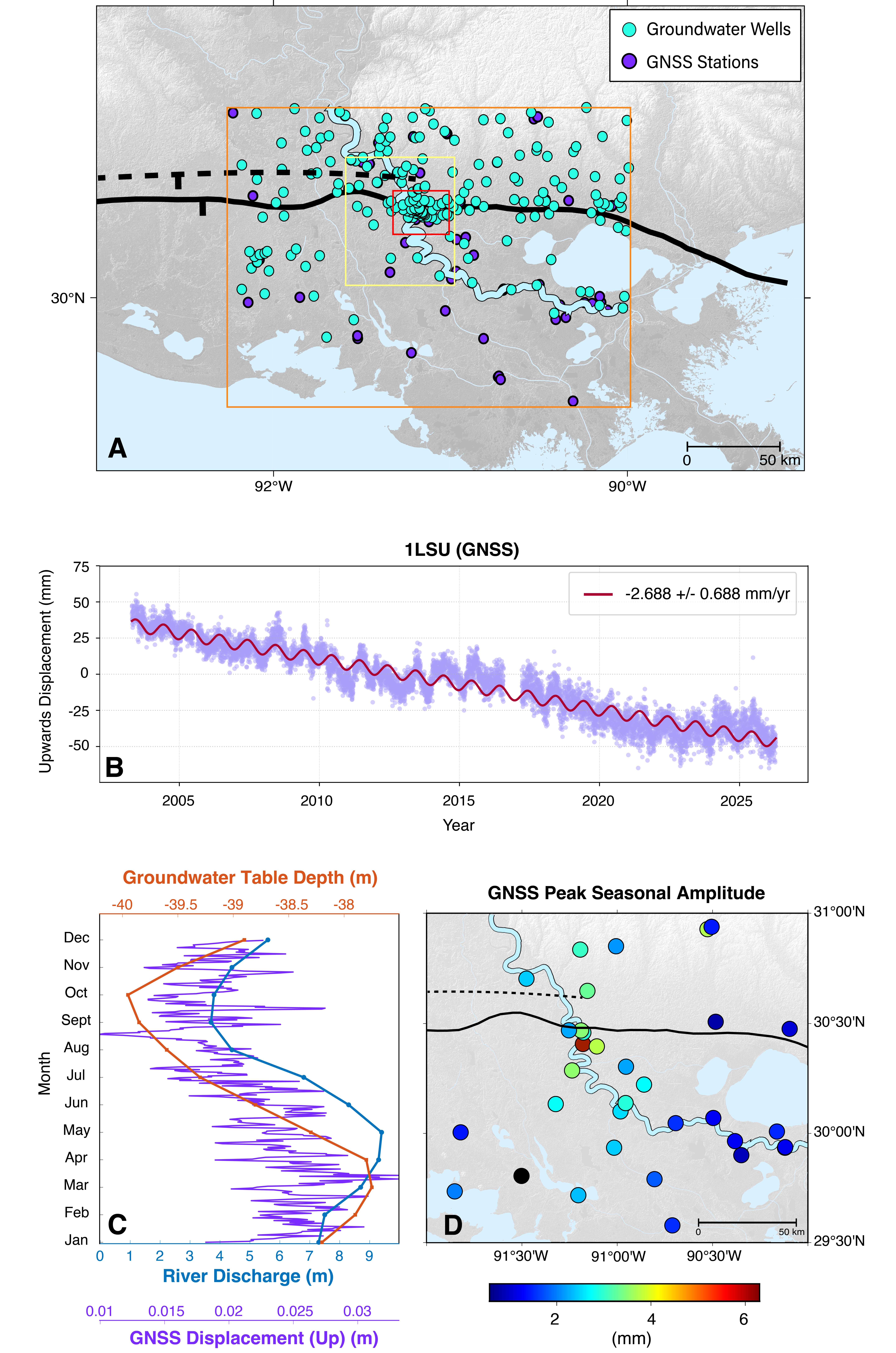}
\caption{\textbf{Geodetic and hydrological observations in southern Louisiana.}
(\textbf{A}) Study area: Baton Rouge Fault (solid) and Denham Springs Fault (dashed), GNSS stations (purple), and monitoring wells (cyan). Boxes mark the regional (orange), focus (yellow), and near-source (red) zones.
(\textbf{B}) Daily vertical position at 1LSU, 2004--2024; the red line is the $-2.69\pm0.69$~mm\,yr$^{-1}$ MIDAS trend.
(\textbf{C}) Monthly-stacked seasonal cycle of Mississippi River stage (blue), hydraulic head at Eb-1274 (orange; Eb-778 is in phase), and 1LSU displacement (purple), all peaking during spring high water.
(\textbf{D}) Peak-to-peak seasonal amplitude by station, largest (12--14~mm) near the Baton Rouge Fault--Mississippi River intersection and decaying outward. Only Pleistocene-terrace stations enter the spatial-decay fit; stations on Holocene alluvium are dominated by shallow compaction and show uniformly small amplitudes (Fig.~\ref{fig:geolmap}).}
\label{fig:obs}
\end{figure}

Daily vertical positions at the 1LSU continuous Global Navigation Satellite System (GNSS) station near Baton Rouge, processed at the Nevada Geodetic Laboratory, span 2004--2024 (Fig.~\ref{fig:obs}A,B).
A MIDAS robust trend estimator~\cite{Blewitt2016} yields a long-term vertical velocity of $-2.69\pm0.69$~mm\,yr$^{-1}$, consistent with regional subsidence rates from earlier work in coastal Louisiana~\cite{Dokka2006,karegar2016subsidence, Karegar2020}.
Cumulative subsidence at 1LSU approaches 55~mm over the observation window.

Superposed on this trend is a clear annual oscillation with peak-to-peak amplitude of 10--15~mm at stations near the Mississippi River.
Monthly-stacked time series (Fig.~\ref{fig:obs}C) reveal that GNSS vertical displacement, river stage at USGS gauge 07374000, and hydraulic head in nearby monitoring wells Eb-778 and Eb-1274, both screened in the deep, confined ``2,000-foot sand'' of the Southern Hills aquifer system (Materials and Methods), are all locked in phase, peaking in February--May and reaching minima in September--November.
GNSS vertical motion lags this hydrological forcing (the seasonal cycle of Mississippi River stage and of confined-aquifer hydraulic head) by 1--2 months, consistent with the time required for pressure diffusion through the confined aquifer system (Materials and Methods provide full data-processing details).
The seasonal forcing itself is ultimately set by the hydroclimatology of the Mississippi River Basin, in which the spring high-water period reflects the seasonal cycle of precipitation and runoff integrated over the drainage area, while the late-summer minima coincide with the seasonal deficit in basin-wide rainfall~\cite{McCabe2019}.

The sign of the correlation is diagnostic: high water levels coincide with ground \textit{uplift}, not subsidence~\cite{Larochelle2022,Carlson2024}.
This positive correlation rules out surface elastic loading as the dominant mechanism and points instead to internal poroelastic loading, in which the aquifer expands as pore pressure rises and contracts as it falls.
The amplitude of the seasonal hydraulic head variation in the deep aquifer (well Eb-1274) is $\dHs \approx 0.5$--1.0~m, while the seasonal river stage variation at USGS gauge 07374000 reaches 5--10~m.
The order-of-magnitude attenuation between river stage and aquifer head, combined with the multi-month phase lag, indicate that the hydraulic connection between the river and the aquifer is mediated rather than direct.

\subsection*{Spatial decay from the fault--river intersection}

The seasonal amplitude $\Delta u$ at GNSS-CORS stations sited on the Pleistocene terrace deposits of southern Louisiana decays systematically with radial distance $r$ from the intersection of the Baton Rouge Fault and the Mississippi River (Fig.~\ref{fig:obs}D).
Stations within 20~km of this intersection show amplitudes of 12--14~mm; stations 80--100~km distant show 4--6~mm.
The pattern is reproducible across years and is robust to alternative trend-removal choices.

The analysis is restricted to stations on the Pleistocene terraces, where laterally continuous sand aquifers confined by compacted clay seals can sustain a confined poroelastic response; the radial decay emerges only there, most sharply at Baton Rouge and more weakly near Lafayette to the west (Fig.~\ref{fig:geolmap}).
Stations on Holocene alluvium and coastal marsh instead overlie thick unconsolidated sequences dominated by shallow sediment compaction~\cite{Tornqvist2008,Jankowski2017,Song2025}; they show uniformly small seasonal amplitudes that do not follow Eq.~\ref{eq:spatial_decay} and are excluded from the fit.
The mechanism is therefore best resolved geodetically where terrace aquifers permit it, and the weak signals elsewhere reflect surface geology rather than the absence of fault-mediated recharge.

To first order the decay is well fit by a logarithmic function of the radial distance $r$ from the intersection of the Mississippi River and the Baton Rouge Fault,
\begin{equation}
	\Delta u(r) = a + b \ln\!\left(r\sqrt{\frac{\omega}{2\alpha_r}}\right),
	\label{eq:spatial_decay}
\end{equation}
where $\omega = 2\pi/$year is the annual angular frequency and $\alpha_r$ is the radial hydraulic diffusivity of the aquifer system.
The best-fit value of the decay parameter is $\sqrt{\omega/(2\alpha_r)} = 5.1\times10^{-6}$~m$^{-1}$.
Values of this parameter for unconsolidated to semi-consolidated sand-and-gravel aquifer systems span roughly $10^{-6}$--$10^{-4}$~m$^{-1}$~\cite{Freeze1979,chen2023unstructured,kuang2020review}, so the fitted decay lies within the range observed in natural aquifers.
We report $\sqrt{\omega/(2\alpha_r)}$ rather than $\alpha_r$ itself because it is the quantity the fit constrains directly; note that changing the constant inside the logarithm from $\sqrt{\omega/(2\alpha_r)}$ to the $\sqrt{\omega/(4\alpha_r)}$ of the analytical solution (Eq.~\ref{eq:neumann}) shifts only the intercept $a$ by $\tfrac{1}{2}\ln 2$ and leaves the fitted slope $b$ unchanged.
The localized, radially decaying pattern centered on the fault--river crossing leads to a key inference: \textit{the fault zone acts as a seasonally varying fluid conduit between the river and the deep aquifer system}, providing enhanced vertical permeability that connects surface and subsurface hydrology~\cite{Bense2006,Bense2013}.

\subsection*{Poroelastic model and analytical framework}

\begin{figure}[htbp]
\centering
\includegraphics[width=\textwidth]{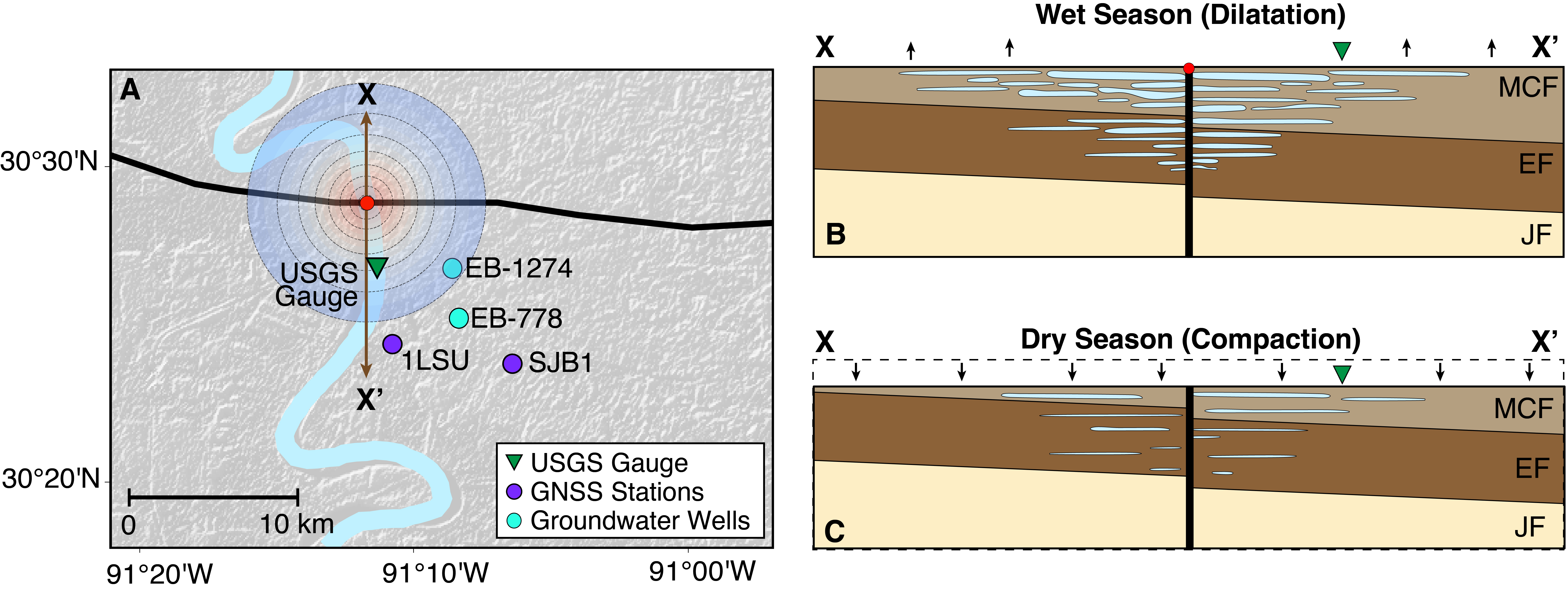}
\caption{\textbf{Conceptual model of fault-mediated seasonal poroelastic loading.}
(\textbf{A}) Map view of the Mississippi River crossing the Baton Rouge Fault (black). The fault--river intersection (red dot) acts as a localized source; dashed rings are radial pressure-diffusion fronts and X--X$'$ locates the cross-sections. Also shown are the USGS river gauge (green triangle), wells Eb-778 and Eb-1274 (cyan), and GNSS stations 1LSU and SJB1 (purple).
(\textbf{B}) Dry season. At low river stage, reduced head at the source lowers pore pressure in the deep confined aquifers; the aquifer contracts and the surface subsides (downward arrows).
(\textbf{C}) Wet season. At high river stage, elevated head raises pore pressure, transmitted along the high-permeability fault rather than by bulk through-fault flow; the aquifer expands and the surface uplifts (upward arrows).
Schematic units of the Southern Hills sand-and-clay sequence~\cite{chen2023unstructured} are the Mississippi and Chicot (MCF), Evangeline (EF), and Jasper (JF) Formations; the most responsive reservoirs, including the heavily pumped ``2,000-foot sand'', lie at 300--800~m depth.}
\label{fig:model}
\end{figure}

We model the aquifer as an oblate spheroidal Eshelby inclusion embedded in an elastic half-space~\cite{Eshelby1957,Segall1998,DavidZimmerman2011} (Fig.~\ref{fig:model}).
The real aquifer geometry, a sequence of interbedded sands and clays from 122 to 823~m depth, is more complex~\cite{Sargent2011}, but this idealization captures the essential coupling between pore-pressure change and volumetric strain in a laterally extensive, vertically thin layer.
For a uniform pressure change $\Delta p$ in an aquifer of thickness $z_d$, the surface vertical displacement is~\cite{Biot1941,Wang2000}
\begin{equation}
	u_z = \frac{(1-2\nu)(2-\nu)}{2(1-\nu^2)\mu}\,\Delta p\,z_d ,
	\label{eq:uz}
\end{equation}
where $\nu$ is Poisson's ratio and $\mu$ the shear modulus.
Equation~\ref{eq:uz} gives the thickness change of the deforming layer; for a source embedded beneath a traction-free surface the free-surface (image) contribution doubles the resulting vertical displacement of the ground, and it is this surface expression that appears in the prefactor of Eqs.~\ref{eq:neumann} and~\ref{eq:longterm} below.
Assuming a hydrostatic pressure response to head changes ($\Delta p = \rho_w g\,\Delta H$), Eq.~\ref{eq:uz} yields scaling relations for both the long-term subsidence and the seasonal amplitude.
Throughout, we distinguish two components of the generic head change $\Delta H$: the seasonal head amplitude $\dHs$ (subscript $s$), which forces the seasonal displacement, and the long-term rate of head decline $\dot H_d$ in the deep confined aquifer (subscript $d$), which forces the secular subsidence.

For the seasonal component, pressure diffuses radially from the fault--river source through a thin laterally extensive aquifer, governed by
\begin{equation}
	\frac{\partial P}{\partial t} = \frac{\alpha_r}{r}\frac{\partial}{\partial r}\!\left(r\frac{\partial P}{\partial r}\right) .
	\label{eq:diffusion}
\end{equation}
We impose on Eq.~\ref{eq:diffusion} a flux (Neumann) boundary condition at the source,
\begin{equation*}
	\left[r\,\frac{\partial P}{\partial r}\right]_{r\to 0} = \Pi\,\mathrm{e}^{i\omega t} ,
\end{equation*}
in which the fault zone, of finite permeability, throttles seasonal fluid exchange between the high-stage river and the deeper aquifer.
Combining the resulting pressure solution with Eq.~\ref{eq:uz} gives the seasonal surface displacement amplitude
\begin{equation}
	\Delta u_{\mathrm{season}}(r) = \frac{(1-2\nu)(2-\nu)\rho_w g z_d}{(1-\nu^2)\mu}\,\dHs\,\ln\!\left(\sqrt{\frac{\omega}{4\alpha_r}}\,r\right) ,
	\label{eq:neumann}
\end{equation}
a logarithmic decay with radial distance (Materials and Methods).
Equation~\ref{eq:neumann} is testable against the observed spatial decay parameterized by Eq.~\ref{eq:spatial_decay}.

The same framework gives the long-term response.
Over the observation window the head in the deep confined aquifer declines at an approximately constant rate $\dot H_d$, so that $\Delta p = \rho_w g\,\dot H_d\,\Delta t$ and Eq.~\ref{eq:uz} predicts a cumulative displacement that grows linearly with elapsed time $\Delta t$,
\begin{equation}
	\Delta u_{\mathrm{long}}(\Delta t) = \frac{(1-2\nu)(2-\nu)\rho_w g z_d}{(1-\nu^2)\mu}\,\dot H_d\,\Delta t ,
	\qquad
	v_z = \frac{\mathrm{d}\,\Delta u_{\mathrm{long}}}{\mathrm{d}t} = \frac{(1-2\nu)(2-\nu)\rho_w g z_d}{(1-\nu^2)\mu}\,\dot H_d .
	\label{eq:longterm}
\end{equation}
Equations~\ref{eq:neumann} and~\ref{eq:longterm} are the seasonal and secular predictions tested below; unlike the seasonal solution, the secular one carries no radial dependence, because the long-term drawdown is imposed by distributed pumping rather than by the fault--river source.

\subsection*{Model--data comparison}

\begin{figure}[htbp]
\centering
\includegraphics[width=\textwidth]{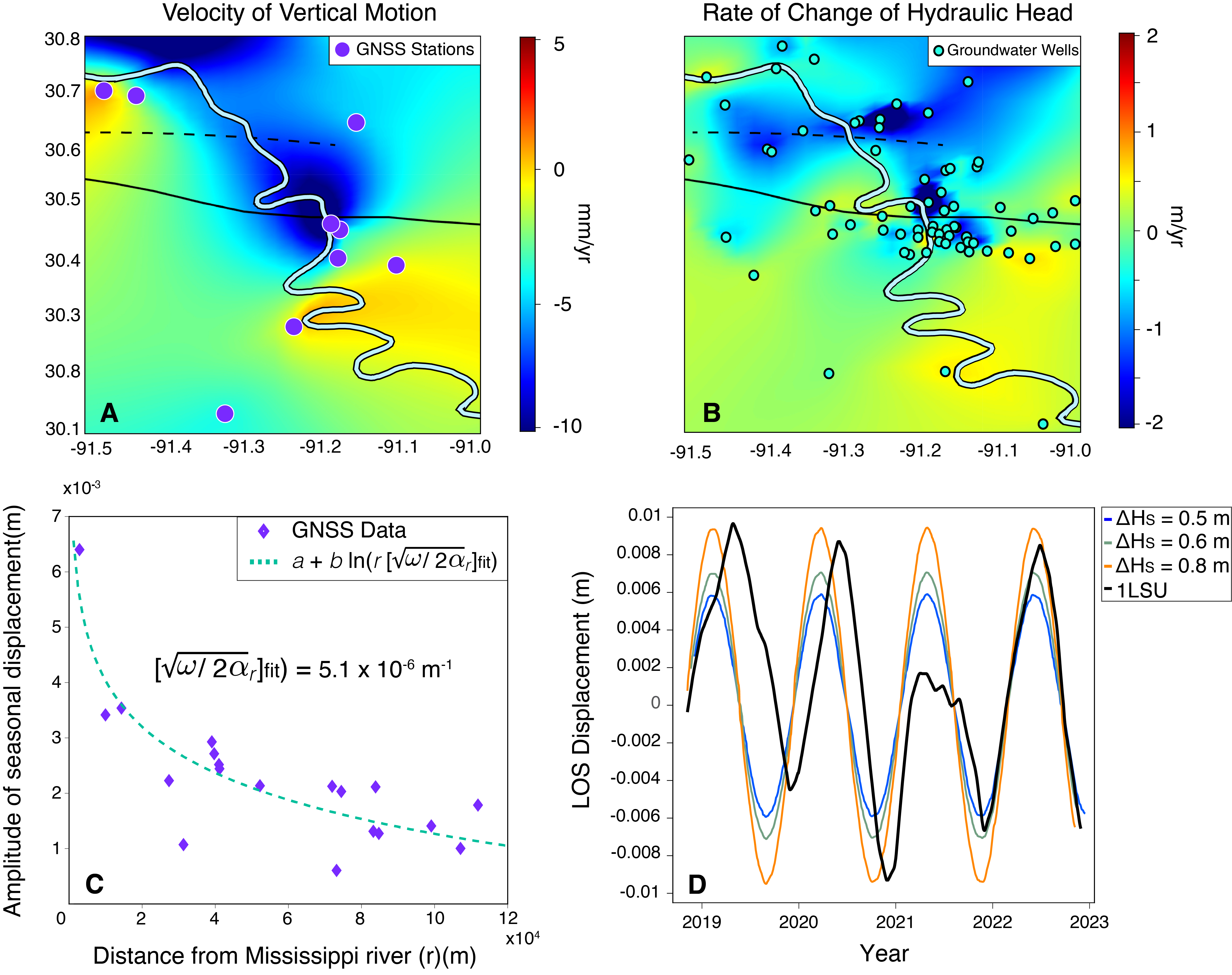}
\caption{\textbf{Validation of the poroelastic model.}
(\textbf{A}) Interpolated long-term GNSS vertical velocity $V_z$ (mm\,yr$^{-1}$) from the focus-area stations shown; subsidence is largest near the fault--river intersection.
(\textbf{B}) Interpolated long-term rate of hydraulic-head change $\dot H_d$ (m\,yr$^{-1}$) in deep-screened wells, the quantity forcing secular subsidence in Eq.~\ref{eq:longterm}. Its spatial correlation with panel A supports a poroelastic origin for the decadal trend.
(\textbf{C}) Observed seasonal amplitude (purple diamonds) versus radial distance $r$ from the intersection, for Pleistocene-terrace stations (Fig.~\ref{fig:geolmap}). The dashed teal line is the fit to Eq.~\ref{eq:neumann}, giving $[\sqrt{\omega/(2\alpha_r)}]_{\text{fit}} = 5.1\times10^{-6}$~m$^{-1}$.
(\textbf{D}) Observed seasonal displacement at 1LSU (black) against predictions for seasonal head amplitudes $\dHs = 0.5$, 0.6, and 0.8~m (blue, green, orange). The 0.6~m case matches the head amplitude measured independently at Eb-1274.}
\label{fig:validation}
\end{figure}

Three independent tests support the poroelastic model.
First, the long-term scaling (Eq.~\ref{eq:longterm}) ties cumulative surface displacement, and hence vertical velocity, to the rate of head change.
The interpolated rate of hydraulic-head decline in wells screened in the deep aquifer is spatially correlated with the interpolated GNSS-derived vertical velocity field (Fig.~\ref{fig:validation}A,B): both reach their largest negative values near the fault--river intersection, where groundwater extraction is most intense~\cite{Sargent2011,Karegar2020,hurtado2026quantifying}.
The observed subsidence rate at 1LSU ($-2.69$~mm\,yr$^{-1}$) is reproduced for $\mu = 5\times10^8$~Pa, $\nu = 0.3$, $z_d = 500$~m and a rate of head decline $\dot H_d \sim -0.4$~m\,yr$^{-1}$, a value consistent with the head-decline rates mapped from the deep-screened monitoring wells (Fig.~\ref{fig:validation}B).

Second, the spatial decay of seasonal amplitude with radial distance from the fault--river intersection is well fit by Eq.~\ref{eq:neumann} (Fig.~\ref{fig:validation}C).
The fitted decay parameter is $\sqrt{\omega/(2\alpha_r)} = 5.1\times10^{-6}$~m$^{-1}$, which falls inside the $10^{-6}$--$10^{-4}$~m$^{-1}$ range spanned by specific-storage and hydraulic-conductivity values reported for unconsolidated to semi-consolidated sand aquifers~\cite{kuang2020review} and by the high-fidelity groundwater model of the Southern Hills aquifer system calibrated against observed heads and withdrawals in southeastern Louisiana~\cite{chen2023unstructured}.
The fit is therefore consistent with independently observed properties of this aquifer system rather than requiring anomalous ones.
The success of the logarithmic decay confirms that the fault zone behaves as a fluid flux source whose magnitude tracks river stage, consistent with a finite-permeability fault providing a throttled hydraulic connection between the river and the deep aquifer.

Third, the seasonal time series at 1LSU is reproduced quantitatively by Eq.~\ref{eq:neumann}.
With aquifer thickness $z_d = 500$~m (equivalently, the depth to the base of the deforming aquifer sequence), radial distance $r = 4$~km, and $\dHs = 0.6$~m (the value independently measured at well Eb-1274), the predicted amplitude matches the observed 10--15~mm peak-to-peak signal (Fig.~\ref{fig:validation}D).
Models with $\dHs = 0.5$ and 0.8~m bracket the observations, demonstrating the model's first-order sensitivity to the imposed head amplitude.

\subsection*{Sensitivity and climate-change implications}

\begin{figure}[htbp]
\centering
\includegraphics[width=\textwidth]{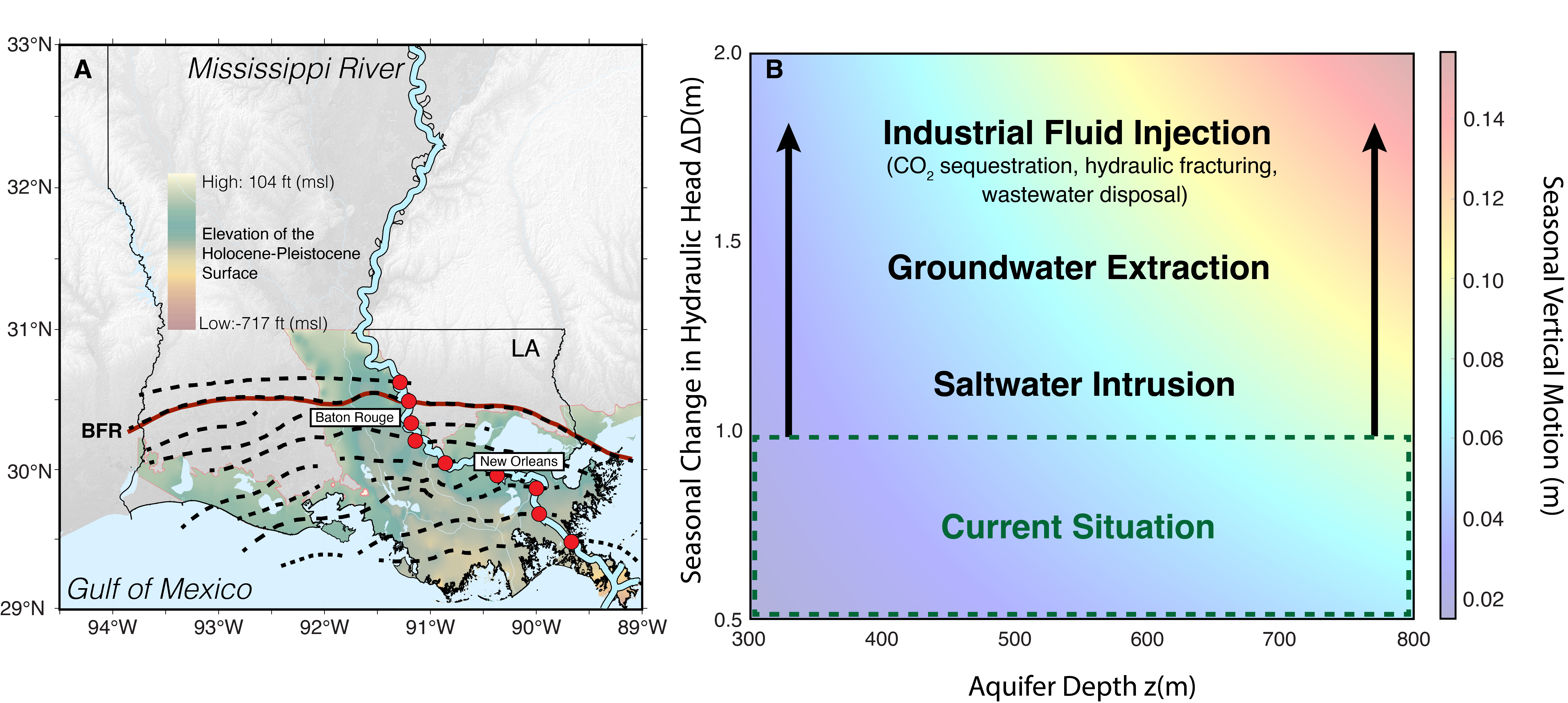}
\caption{\textbf{Regional implications and climate-change projection.}
(\textbf{A}) Coastal Louisiana, showing down-to-the-south Pleistocene growth faults (dashed) and their intersections with the Mississippi River and its distributaries (red dots). The Baton Rouge Fault (BRF) is arrowed; BR is Baton Rouge, NO is New Orleans. Shading gives depth to the buried Holocene--Pleistocene surface, so darker areas carry thicker Holocene cover. Each intersection is a candidate site for the mechanism documented here, but the signal is geodetically resolvable only where confined Pleistocene terrace aquifers lie at or near the surface (Fig.~\ref{fig:geolmap}).
(\textbf{B}) Predicted seasonal vertical displacement (colour) as a function of aquifer thickness $z_d$ and seasonal head amplitude $\dHs$, from Eq.~\ref{eq:neumann}. The dashed box marks present southern Louisiana conditions; arrows and labelled bands mark three anthropogenic trajectories toward larger amplitudes: saltwater intrusion, groundwater extraction, and industrial fluid injection.}
\label{fig:implications}
\end{figure}

Equations~\ref{eq:uz} and~\ref{eq:neumann} are linear in both aquifer thickness $z_d$ and seasonal head amplitude $\dHs$.
Mapping predicted seasonal vertical displacement across the parameter space (Fig.~\ref{fig:implications}B) shows that current Southern Louisiana conditions ($z_d \approx 400$--600~m, $\dHs \approx 0.5$--1.0~m) yield seasonal amplitudes of order centimeters.
Three anthropogenic trajectories push the system toward larger displacements (Fig.~\ref{fig:implications}B).

\textit{Intensified hydrological extremes.} Wetter wet seasons and drier dry seasons in a warming climate~\cite{Taylor2013} raise $\dHs$ directly, and Eq.~\ref{eq:neumann} is linear in that amplitude.

\textit{Groundwater extraction and saltwater intrusion.} Continued withdrawal deepens the cone of depression and sustains $\dot H_d$~\cite{GallowayBurbey2011,Ojha2018}. In the Baton Rouge system it also drives salinization: the freshwater--saltwater interface coincides with the Baton Rouge Fault, and withdrawals from the ``2,000-foot'' sand have produced drawdowns of up to $\sim$108~m and pulled saltwater northward across the fault into the freshwater side~\cite{Heywood2014,tomaszewski2002water}. The effect of that salinization on vertical land motion is indirect, and its sign is worth stating explicitly, because sea-level rise on its own would raise coastal heads and slightly \textit{reduce} drawdown-driven subsidence~\cite{Ferguson2012}. The operative pathway here runs the other way: salinization removes usable storage from the shallower sands and forces withdrawal to be redistributed into deeper and more restricted intervals, which sustains $\dot H_d$ and increases the thickness $z_d$ of the poroelastically responding column. Both changes increase the secular and the seasonal displacement predicted by Eqs.~\ref{eq:neumann} and~\ref{eq:longterm}. Separating this indirect pathway from the direct effect of extraction will require geodetic and chloride records to be analysed jointly and is beyond the scope of this study.

\textit{Industrial fluid injection.} Wastewater disposal, hydraulic fracturing and, prospectively, large-scale CO\textsubscript{2} storage all raise pore pressure in the same deep, laterally extensive saline aquifers that respond poroelastically to seasonal forcing, and these formations are leading candidates for Gulf Coast CO\textsubscript{2} storage~\cite{Meckel2021,Bump2021,Callas2022,ZobackHennings2025}. An injection-driven pore-pressure increase would be superimposed on the natural seasonal cycle documented here and would add to the volumetric strain budget; because the fault provides the hydraulic connection that produces the seasonal signal in the first place, pressure-management strategies for storage operations need to account for this pre-existing, fault-mediated pressure signal rather than treating the reservoir as hydraulically isolated~\cite{BumpHovorka2024,Zulqarnain2023}.

For $\dHs > 1.5$~m the predicted seasonal amplitude can grow 2--3 fold over present values, plausibly reaching 10~cm or more in areas overlying the thickest aquifers.

\subsection*{Broader significance}

Four implications follow.
\textit{First}, poroelastic deformation dominates the seasonal vertical signal in Southern Louisiana, a result that likely extends to other deltaic settings where thick permeable aquifers are hydraulically connected to large rivers.
Geodetic time series in such regions cannot be detrended of hydrology by assuming surface-load anti-correlation; doing so will bias both secular subsidence estimates and the attribution of residual signals to tectonic processes~\cite{Amelung1999,Chaussard2014}.
\textit{Second}, the fault--river intersection acts as a discrete fluid source rather than a distributed recharge zone.
Multiple sub-parallel Pleistocene growth faults intersect the Mississippi River and its distributaries across coastal Louisiana (Fig.~\ref{fig:implications}A); each is a candidate site for the same mechanism, implying a spatially patchy contribution to regional subsidence that must be mapped fault-by-fault for accurate hazard assessment~\cite{Allison2016,Shirzaei2021,Jones2016,Minderhoud2018}.
This signature is expressed geodetically only where confined terrace aquifers are present, so GNSS resolves the mechanism where the surface geology permits rather than wherever it operates: fault-mediated recharge plausibly occurs more widely but is masked at stations on compacting Holocene deposits, where shallow compaction dominates the seasonal signal~\cite{Song2025}.
\textit{Third}, the poroelastic component of vertical land motion is sensitive to hydrological forcing through the linear dependence of Eq.~\ref{eq:neumann} on $\dHs$.
As precipitation variability and extreme river-stage events grow under climate change, the seasonal amplitude of poroelastic ground motion will grow proportionally.
This previously underappreciated coupling (between hydrological extremes and crustal deformation through fault-mediated aquifer recharge) represents a quantifiable, model-based bridge between climate projections and coastal-subsidence forecasts.

\textit{Fourth}, our results bear directly on plans for large-scale geologic storage of CO\textsubscript{2} along the Gulf Coast.
The geologically young, diagenetically immature, and weakly cemented sedimentary formations of coastal Texas and Louisiana have been repeatedly identified as among the most favorable settings in North America for carbon storage, both in saline aquifers and in depleted hydrocarbon reservoirs, and have become the focus of an expanding portfolio of proposed storage hubs~\cite{Meckel2021,Bump2021,DeAngelo2019,Zulqarnain2023}.
The case for the region rests in large part on its pervasive clay-rich seals and on growth faults that are widely assumed to slip aseismically, limiting the magnitude of any triggered seismicity~\cite{ZobackHennings2025}.
Yet these favorability assessments have been built almost exclusively from static geological and geomechanical characterization (seal continuity, structural mapping, pore volume, and in situ stress) and have not accounted for the dynamic, fault-mediated hydraulic connectivity that our observations reveal.
We show that at least one major growth fault transmits a measurable seasonal pressure signal between surface water and a deep aquifer, on time scales of months and over distances of tens of kilometers.
A fault capable of throttling seasonal fluid exchange in this way is, by the same token, a potential pathway for the vertical migration of pressure, and ultimately of buoyant CO\textsubscript{2}, out of an intended storage interval.
The seasonal poroelastic signal documented here thus provides an observable, model-based diagnostic of which faults are hydraulically active under natural forcing, information that is presently absent from Gulf Coast site-selection workflows but that bears directly on long-term storage security and on the pressure-management strategies that those operations will require~\cite{BumpHovorka2024,Callas2022}.
Incorporating fault--aquifer connectivity of the kind we identify into storage-site characterization would close a significant gap between the current assessment of the region as ``good'' for sequestration and the dynamic subsurface behavior that controls its actual performance.

% Length check (draft 8): main text 2,701 words (limit 3,000); abstract 124 (limit 125);
% title 87 chars (limit 96); 4 main display items (limit 5); 59 references (~50 guideline).

%%%%%%%%%%%%%%%% REFERENCES %%%%%%%%%%%%%%%

\clearpage

\bibliography{poroelastic_louisiana}
\bibliographystyle{sciencemag}

%%%%%%%%%%%%%%%% ACKNOWLEDGEMENTS %%%%%%%%%%%%%%%

\paragraph*{Funding:}
P.S., E.~Aharonov and C.E. acknowledge the support of the Bi-national Israel--US Industrial Development Fund of the US--Israel Energy Center for Fossil Energy. E.~Arzabala acknowledges support from the Gates Millennial Foundation. C.E. acknowledges support from the Marshall-Heape endowment to Tulane. R.T. acknowledges the support of the University of Oslo, the Njord Centre, the CNRS IRP D-FFRACT and the Research Council of Norway through the PoreLab Center of Excellence (project number 262644). S.P. acknowledges the support of the Johannes Amos Comenius Programme (P~JAC), project No.~CZ.02.01.01/00/22\_008/0004605, Natural and Anthropogenic Georisks.
\paragraph*{Author contributions:}
P.S. led the analysis and wrote the manuscript. E.~Arzabala assisted with data analysis and data visualization. C.H.-P. and C.E contributed data analysis and the regional geological synthesis. E.~Aharonov, R.T., and S.P. contributed to the theoretical framework and analytical derivations. C.E. designed and supervised the project. All authors contributed to interpretation and manuscript revision.
\paragraph*{Competing interests:}
There are no competing interests to declare.
\paragraph*{Data and materials availability:}
GNSS time series are publicly available from the Nevada Geodetic Laboratory (\url{http://geodesy.unr.edu/}).
Groundwater levels and river-stage data are publicly available from the U.S.~Geological Survey National Water Information System (\url{https://waterdata.usgs.gov/nwis}).
Processed amplitude/velocity tables used in this study are archived at a public repository~\cite{sarma2026dataset}.

%%%%%%%%%%%%%%%% SUPPLEMENT LIST %%%%%%%%%%%%%%%

\subsection*{List of Supplementary Materials}
Materials and Methods\\
Fig. S1\\
Table S1\\
References \textit{(56--59)} % Blewitt2018, OleaColeman2014, AbramowitzStegun1964, Tadeu2004 are cited only in the supplement

%%%%%%%%%%%%%%%% END OF MAIN TEXT %%%%%%%%%%%%%%%

\newpage

%%%%%%%%%%%%%%%% START OF SUPPLEMENT %%%%%%%%%%%%%%%

% Figures, tables, equations and pages in the supplement are numbered S1, S2 etc.
\renewcommand{\thefigure}{S\arabic{figure}}
\renewcommand{\thetable}{S\arabic{table}}
\renewcommand{\theequation}{S\arabic{equation}}
\renewcommand{\thepage}{S\arabic{page}}
\setcounter{figure}{0}
\setcounter{table}{0}
\setcounter{equation}{0}
\setcounter{page}{1}
% References continue the numbering from the main text.

%%%%%%%%%%%%%%%% SUPPLEMENT TITLE PAGE %%%%%%%%%%%%%%%

\begin{center}
\section*{Supplementary Materials for\\ \scititle}

Pritom~Sarma$^{\ast}$,
Eduardo~Arzabala,
Carolina~Hurtado-Pulido,
Einat~Aharonov,
Renaud~Toussaint,
Stanislav~Parez,
Cynthia~Ebinger\\
\small$^\ast$Corresponding author. Email: pritom.sarma@mail.huji.ac.il
\end{center}

\subsubsection*{This PDF file includes:}
Materials and Methods\\
Fig. S1\\
Table S1

\newpage

%%%%%%%%%%%%%%%% MATERIALS AND METHODS %%%%%%%%%%%%%%%

\subsection*{Materials and Methods}

\subsubsection*{GNSS data and processing}

Continuous Global Navigation Satellite System (GNSS) daily position time series were obtained from the Nevada Geodetic Laboratory (NGL) for all available GNSS-CORS stations within the regional study area (latitudes 29\textdegree~N--31.5\textdegree~N, longitudes 89\textdegree~W--93\textdegree~W).
NGL provides daily solutions in the IGS14 reference frame, processed with the GIPSY-OASIS software using precise point positioning~\cite{Blewitt2018}.

\begin{figure}[htbp]
\centering
\includegraphics[width=\textwidth]{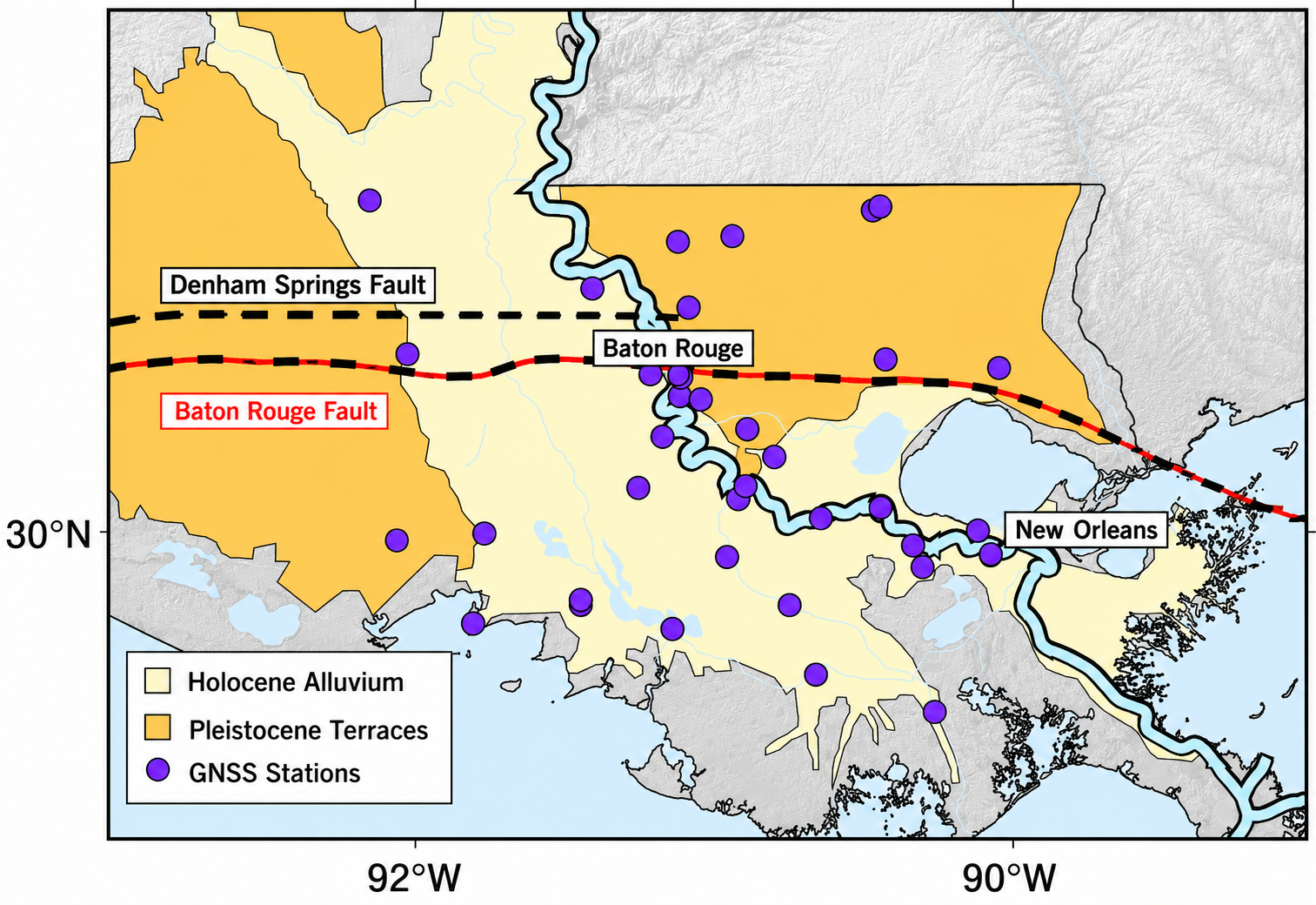}
\caption{\textbf{Regional geologic setting and lithologic basis for station selection.}
Generalized geologic map of southern Louisiana, modified after~\cite{OleaColeman2014} (base map: Louisiana Geological Survey, revised 2010), showing the surface distribution of Pleistocene terrace deposits (orange) and Holocene alluvium (pale yellow), together with the Baton Rouge Fault (red), the Denham Springs Fault (black dashed), the Mississippi River, the cities of Baton Rouge and New Orleans, and the continuous GNSS stations used here (purple).
The confined sand aquifers that carry the seasonal poroelastic signal are developed within the Pleistocene terraces; the spatial-decay analysis of the main text (Figs.~\ref{fig:obs}D and~\ref{fig:validation}C) is therefore restricted to stations sited on these terraces.
Stations overlying Holocene alluvium are dominated by shallow sediment compaction, and register incoherent seasonal poroelastic response and present a misfit in Fig~3C ~\cite{Tornqvist2008,Jankowski2017,Song2025}.}
\label{fig:geolmap}
\end{figure}

Quality control retained only daily solutions with formal vertical uncertainties below 10~mm, and only stations with at least three years of continuous data.
For the spatial-decay analysis (Figs.~\ref{fig:obs}D and~\ref{fig:validation}C of the main text), the station set was further restricted to sites located on the Pleistocene terrace deposits, where laterally continuous sand aquifers separated by clay seals are developed; stations overlying Holocene alluvium and coastal-marsh deposits, whose seasonal vertical motion is dominated by shallow sediment compaction, were excluded (Fig.~\ref{fig:geolmap})~\cite{Tornqvist2008,Jankowski2017,Song2025}.
Long-term vertical velocities at each station were estimated with the MIDAS (Median Interannual Difference Adjusted for Skewness) algorithm~\cite{Blewitt2016}, which is robust to step discontinuities and seasonal variability.
The interpolated velocity and head-rate maps (Figs.~\ref{fig:validation}A,B) were generated from the GNSS stations and monitoring wells located within the focus area shown in those panels, rather than from the full regional network of Fig.~\ref{fig:obs}A.

Seasonal amplitudes were extracted by stacking the residual time series (daily positions minus the MIDAS linear trend and any documented station offsets) by calendar month over the full 2004--2024 window.
This procedure suppresses interannual variability while preserving the periodic annual signal.
The peak-to-peak seasonal amplitude at each station was taken as the maximum minus the minimum monthly-stacked vertical position.
Uncertainties in seasonal amplitude were estimated by bootstrap resampling of years (1000 replicates).

\subsubsection*{Groundwater and river-stage data}

Hydraulic head data were retrieved from the U.S.~Geological Survey National Water Information System (NWIS) for monitoring wells within the study area.
Only wells screened below 100~m depth were retained, so that the analysed head records sample the confined aquifers that carry the poroelastic signal rather than the shallow water table; the two wells used in the time-series analysis, Eb-778 and Eb-1274, are screened in the so-called ``2,000-foot sand'' near Baton Rouge.

River-stage data are daily mean gauge-height values from USGS stream gauge 07374000 (Mississippi River at Baton Rouge), retrieved from NWIS over 2004--2024.
Both groundwater and river-stage time series were stacked by calendar month using the same procedure as for GNSS.

\subsubsection*{Cross-correlation analysis}

The lag between hydrological forcing and surface displacement was estimated by computing the cross-correlation between (i) monthly-resampled GNSS vertical residuals at 1LSU and (ii) hydraulic head at Eb-1274 and river stage at gauge 07374000.
Both signals were detrended and tapered before cross-correlation.
The peak cross-correlation occurs at a positive lag of 1--2 months, with the geodetic signal trailing the hydrological forcing.

This lag is consistent with the time required for pressure to equilibrate vertically through the confined sequence. For one-dimensional diffusion over a thickness $z_d$ the characteristic time is
\begin{equation}
	t_{\mathrm{diff}} \simeq \frac{z_d^{2}}{4\alpha_z} ,
	\label{eq:tdiff}
\end{equation}
where $\alpha_z$ is the vertical hydraulic diffusivity. Taking $z_d \approx 500$~m, a lag of 1--2 months corresponds to $\alpha_z \approx 4\times10^{5}$--$8\times10^{5}$~m$^{2}$\,yr$^{-1}$, within the $10^{4}$--$10^{6}$~m$^{2}$\,yr$^{-1}$ range reported for unconsolidated to semi-consolidated sand aquifers~\cite{Freeze1979,kuang2020review}. The observed phase lag is therefore reproduced without adjusting the aquifer properties away from independently measured values.

\subsubsection*{Analytical poroelastic framework}

All symbols and parameter values used in this section are defined in Table~\ref{tab:params}.

\paragraph{Governing equation.}
We model the seasonal pressure perturbation in a thin, laterally extensive aquifer of vertical thickness $z_d$ and far greater radial extent.
Under the assumption $r \gg z_d$, vertical pressure diffusion equilibrates rapidly compared with radial diffusion (Eq.~\ref{eq:tdiff}), so the three-dimensional diffusion equation reduces to the axisymmetric radial form given as Eq.~\ref{eq:diffusion} of the main text, restated here for reference:
\begin{equation}
	\frac{\partial P}{\partial t} = \frac{\alpha_r}{r}\frac{\partial}{\partial r}\!\left(r\frac{\partial P}{\partial r}\right) .
	\label{eq:diffusion_sm}
\end{equation}
Equation~\ref{eq:diffusion_sm} is the equation to which the boundary condition below is applied. It is solved on $0 < r < \infty$ subject to the far-field condition $P(r\to\infty) \to P_0$ and to a flux condition at the fault--river source, $r\to 0$, specified below.

\paragraph{Separation of variables.}
Because the forcing is periodic at the annual frequency $\omega$, we seek complex-valued solutions
\begin{equation*}
	P(r,t) = P_0 + p(r)\,\mathrm{e}^{i\omega t} ,
\end{equation*}
of which the physical pressure is the real part. Substituting into Eq.~\ref{eq:diffusion_sm} and dividing by $\alpha_r\,p(r)\,\mathrm{e}^{i\omega t}$ separates the variables and gives, for all $r>0$,
\begin{equation}
	r\,p''(r) + p'(r) - \frac{i\omega}{\alpha_r}\, r\, p(r) = 0 .
	\label{eq:bessel_ode}
\end{equation}
Equation~\ref{eq:bessel_ode} is the modified Bessel equation of order zero in the complex argument $q r$, with
\begin{equation}
	q \equiv \sqrt{\frac{i\omega}{\alpha_r}} = \sqrt{\frac{\omega}{2\alpha_r}}\,(1+i), \qquad |q| = \sqrt{\frac{\omega}{\alpha_r}} .
	\label{eq:qdef}
\end{equation}
Its general solution is
\begin{equation}
	p(r) = A\,K_0(q r) + B\,I_0(q r) ,
	\label{eq:general_sol}
\end{equation}
where $I_0$ and $K_0$ are modified Bessel functions of the first and second kind~\cite{AbramowitzStegun1964}. Since $I_0(qr)$ diverges as $r\to\infty$ while $K_0(qr)\to 0$, the far-field condition requires $B=0$, leaving
\begin{equation}
	p(r) = A\,K_0(q r) .
	\label{eq:K0_sol}
\end{equation}
The remaining constant $A$ is fixed by the condition at the source.

\paragraph{Flux (Neumann) condition.}
The fault zone has finite permeability and therefore delivers a bounded seasonal fluid flux to the aquifer rather than imposing the river stage on it directly. Prescribing that flux,
\begin{equation}
	\left[r\,\frac{\partial p}{\partial r}\right]_{r\to 0} = \Pi ,
	\label{eq:neumann_bc}
\end{equation}
and using $K_0'(x) = -K_1(x)$ together with the limit $x K_1(x)\to 1$ as $x\to 0$~\cite{AbramowitzStegun1964}, we obtain
\begin{equation*}
	r\,\frac{\partial}{\partial r}\Big[A K_0(qr)\Big] = -A\,(qr)\,K_1(qr) \;\xrightarrow[r\to 0]{}\; -A ,
\end{equation*}
so that $A = -\Pi$, independent of any source radius. The solution is therefore unique:
\begin{equation}
	p(r) = -\Pi\,K_0(q r) .
	\label{eq:neumann_sol}
\end{equation}

\paragraph{Small-argument (near-source) limit.}
For $|q| r \ll 1$, that is $r \ll \sqrt{\alpha_r/\omega}$, the small-argument expansion of the modified Bessel function is $K_0(x) \simeq -\ln(x/2) - \gamma$, with $\gamma$ the Euler--Mascheroni constant~\cite{AbramowitzStegun1964}. Using $|q|/2 = \sqrt{\omega/(4\alpha_r)}$ from Eq.~\ref{eq:qdef}, the modulus of the pressure amplitude follows the logarithmic form
\begin{equation}
	K_0(q r) \simeq -\ln\!\left(\sqrt{\frac{\omega}{4\alpha_r}}\, r\right) + \text{const.},
	\qquad
	|p(r)| \simeq \Pi\,\ln\!\left(\sqrt{\frac{\omega}{4\alpha_r}}\, r\right) + \text{const.}
	\label{eq:bessel_smallarg}
\end{equation}
The additive constants collect $\gamma$ and the phase of $q$; they shift the intercept of the decay but not its slope, which is why the empirical fit of Eq.~\ref{eq:spatial_decay} can equivalently be written with $\sqrt{\omega/(2\alpha_r)}$ inside the logarithm.

\paragraph{Cross-check by the Green's-function route.}
The same result is obtained by treating the fault--river intersection as a two-dimensional oscillating line source and solving Eq.~\ref{eq:diffusion_sm} by Fourier transform, which yields a Hankel function of the second kind, $H_0^{(2)}\!\big(\sqrt{-i\omega/\alpha_r}\,r\big)$, of the same complex argument~\cite{Tadeu2004}. Its leading small-argument behaviour, $H_0^{(2)}(x) \simeq -(2i/\pi)\ln(x/2)$, reproduces Eq.~\ref{eq:bessel_smallarg} up to a multiplicative constant and a constant phase, confirming the logarithmic near-source decay independently of the separation-of-variables argument.

\paragraph{From pressure to surface displacement.}
The source strength $\Pi$ is set by the seasonal fluid exchange across the fault zone. Writing it in terms of the equivalent seasonal head amplitude that drives the aquifer, $\Pi = 2\rho_w g\,\dHs$, and substituting Eq.~\ref{eq:bessel_smallarg} into Eq.~\ref{eq:uz} of the main text gives the seasonal surface displacement amplitude
\begin{equation}
	\Delta u_{\mathrm{season}}(r) = \frac{(1-2\nu)(2-\nu)\rho_w g\, z_d}{(1-\nu^2)\mu}\,\dHs\,\ln\!\left(\sqrt{\frac{\omega}{4\alpha_r}}\,r\right) ,
	\label{eq:neumann_sm}
\end{equation}
which is Eq.~\ref{eq:neumann} of the main text. The factor of two in the relation between $\Pi$ and $\dHs$ is the free-surface contribution noted after Eq.~\ref{eq:uz}; it rescales the amplitude but leaves the logarithmic spatial decay, and therefore the fitted value of $\sqrt{\omega/(2\alpha_r)}$, unchanged. Substituting instead the secular pressure change $\Delta p = \rho_w g\,\dot H_d \Delta t$, which is spatially distributed rather than sourced at the fault, gives the long-term result quoted as Eq.~\ref{eq:longterm} of the main text.

\subsubsection*{Model parameters}

Best-fit and bracketing model parameters are summarized in Table~\ref{tab:params}. Values are chosen within published ranges for unconsolidated to semi-consolidated sand aquifers~\cite{Freeze1979,Wang2000} and the Southern Hills aquifer system in particular~\cite{chen2023unstructured,tomaszewski2002water,Sargent2011}.

%%%%%%%%%%%%%%%% SUPPLEMENTARY TABLES %%%%%%%%%%%%%%%

\begin{table}
\centering
\caption{\textbf{Model parameters used in the poroelastic analysis.}
Best-fit values reproduce the observed long-term subsidence and seasonal amplitude at the 1LSU station; bracketing ranges reflect uncertainties in aquifer mechanical and hydrological properties documented in the cited literature.}
\label{tab:params}

\begin{tabular}{lccc}
\\
\hline
Parameter & Symbol & Range & Best fit\\
\hline
Shear modulus (Pa)              & $\mu$       & $10^8$--$10^9$       & $5\times10^8$\\
Poisson's ratio                 & $\nu$       & 0.25--0.35           & 0.30\\
Aquifer thickness (m)           & $z_d$       & 300--800             & 500\\
Seasonal head amplitude (m)     & $\dHs$      & 0.5--1.0             & 0.6\\
Radial decay parameter (m$^{-1}$) & $\sqrt{\omega/(2\alpha_r)}$ & $10^{-6}$--$10^{-4}$ & $5.1\times10^{-6}$\\
Source distance (km)            & $r$         & 1--100               & 4\\
Long-term head-decline rate (m\,yr$^{-1}$) & $\dot H_d$ & $-1$ to $-0.2$ & $-0.4$\\
Water density (kg\,m$^{-3}$)    & $\rho_w$    & n/a                  & 1000\\
Gravitational acceleration (m\,s$^{-2}$) & $g$ & n/a                 & 9.81\\
\hline
\end{tabular}
\end{table}

%%%%%%%%%%%%%%%% SUPPLEMENTARY REFERENCES %%%%%%%%%%%%%%%

% Do NOT include a separate reference list in the supplement.
% All references are in a single list at the end of the main text.

\end{document}